\documentclass[twocolumn,aps,amsmath]{revtex4}  
\usepackage{graphicx,natbib}
\newcommand{\erf}{\ensuremath{\mbox{erf}}}

\begin{document}

\title{Depletion forces near a soft surface}
\author{Thomas Bickel}
\email[E-mail address: ]{bickel@physics.ucla.edu}
\thanks{Accepted for publication in The Journal of Chemical Physics.}
\affiliation{Department of Physics and Astronomy, UCLA, Box 951547, Los Angeles CA 90095-1547, USA}
\date{February 28, 2003}

\begin{abstract}

\noindent We investigate excluded-volume effects in a bidisperse colloidal suspension near a flexible interface. Inspired by a recent experiment by Dinsmore \emph{et al.} (Phys. Rev, Lett. {\bf 80}, 409 (1998)), we study the adsorption of a mesoscopic bead on the surface and show that depletion forces could in principle lead to particle encapsulation. We then consider the effect of surface fluctuations on the depletion potential itself and construct the density profile of a polymer solution near a soft interface. Surprisingly we find that the chains accumulate at the wall, whereas the density displays a deficit of particles at distances larger than the surface roughness. This non-monotonic behavior demonstrates that surface fluctuations can have major repercusions on the properties of a colloidal solution. On average, the additional contribution to the Gibbs adsorbance is negative. The amplitude of the depletion potential between a mesoscopic bead and the surface increases accordingly.

\end{abstract}

\maketitle

\section{Introduction}

In a recent experiment, Dinsmore and collaborators~\cite{dinsmorePRL98} measured the probability distribution of a single colloidal bead enclosed in a rigid, pear-shaped vesicle. They find that upon addition of much smaller spheres, density gradients give rise to an effective force field that pushes the particle towards the wall. These depletion forces have been discussed for almost fifty years~\cite{aoJCP54,richmondCPL74,vrijPAC76,joannyJpolSci79}, especially since they can be used to organize self-assembled structures~\cite{xiaAdvMatt00}. The remarkable feature observed in~\cite{dinsmorePRL98} is that entropic interactions drastically affect the distribution in a way that  depends on the \emph{local shape} of the membrane. Indeed, it has been shown~\cite{rothPRL99} that inhomogeneities in the curvature distribution induce additional forces that bring the particle to \emph{inwardly} curved regions. 
The authors then suggested that if curvature induces forces on particles, the latter may in turn lead to shape transitions provided that the membrane is flexible enough~\cite{dinsmorePRL98}.

If this last statement is valid, it would be conceivable to take advantage of depletion forces to mimic biological processes such as protein integration, membrane fusion, or encapsulation. For instance, polyethylene glycol is a (polymeric) depleting agent known for its low toxicity: By changing its mass or concentration, one can selectively tune the range and the depth of the depletion potential without affecting any physiological parameter, i.e., pH, temperature and ionic strengths. This remarkable versatility further justifies the growing interest in depletion interactions and makes them particularly suitable for applications such as gene transfection~\cite{behrPNAS95}.

In this article, we present a systematic study of entropic forces near a fluctuating interface. The additional degree of freedom associated with surface softness induces numerous antagonistic features that have to be carefully accounted for. Besides the eventuality of shape transformation, a fluid membrane experiences thermal undulations whose amplitude is inversely proportional to its bending rigidity~\cite{helfrichZfP73}. The steric forces between adjacent bilayers, originating from shape excitations, are invoked to explain the phase behavior of lamellar stacks of membranes~\cite{helfrichZfP78}. Similarly, a colloidal particle has first to overcome a repulsive barrier prior to adhering to a fluctuating membrane. This point is usually ignored in discussions concerning colloidal adsorption, even though it may deeply affect the kinetics of the binding process. A more refined consequence is the effect of membrane fluctuations on the concentration profile of small particles, which in turn affects the depletion potential. It is the aim of this work to theoretically address these different issues. The paper is organized as follows: In Section~\ref{sec2}, we give a brief outlook of some general results regarding effective, entropic interactions. The balance between depletion and bending forces is studied in Section~\ref{sec3}, where we derive a tentative phase diagram of the system. We present in Section~\ref{sec4} our findings concerning the corrections to the depletion potential. We conclude this report with a discussion and summary of our results in Section~\ref{sec5}.

\section{Effective interactions in a colloidal solution \label{sec2}}

We consider a binary mixture of small and large spherical beads with radius $R_s$ and $R$, respectively. In a single-component description, the large particles can then be seen as moving in an \emph{effective force field} created by the other species~\cite{likosPR01}. From a theoretical point of view, this corresponds to integrate out the partition function over the degrees of freedom of the small particles.  The partial trace can be performed for a dilute solution within the formalism of density functional theory~\cite{rothPRE00}. At low density, the effective potential $W({\bf r})$ acting on a \emph{single} colloid immersed in a ``sea'' of small particles is~\cite{gotzelmannEPL99} 
\begin{equation}
\label{depletion}
\beta W({\bf r}) = -\int d^3 {\bf r'} \left\{ \rho({\bf r'}) -\rho_b \right\} f({\bf r-r'})   \ ,
\end{equation}
with $\beta=(k_BT)^{-1}$, $\rho({\bf r})$ the density of small components and $\rho_b=\rho(\infty)$ the bulk value. The Mayer function equals $f({\bf r})=\exp[-\beta v({\bf r})] -1$, with $v({\bf r})$ the direct interaction potential between small and large components. Note that in Eq.~(\ref{depletion}), the thermodynamic quantities have to be evaluated \emph{before} insertion of the large particle. This assertion happens to be essential since the symmetries of the relevant functions are solely determined by the geometry of the system. For instance in the presence of a planar wall at $z=0$, the density profiles depend on the $z$-coordinate only, whereas translational symmetry in the $(x,y)$-plane is broken after particle insertion. 

Eq.~(\ref{depletion}) provides a low-density expansion for the depletion potential even if the interactions between the species are soft and possibly contains attractive contributions. It also encompasses the hard-sphere limit considered in the original theory of Asakura and Oosawa~\cite{aoJCP54}, for which the Mayer function reduces to $f_{HS}({\bf r})=-H( R+R_s -\vert {\bf r} \vert )$. $H(x)$ is the usual Heaviside step function that equals  $1$ for $x>0$ and $0$ otherwise. 
In the presence of a planar wall, the volume fraction of depleting particles $\varphi(z)\hat{=} 4\pi R_s^3 \rho(z)/3$ is also expressed in terms of the step function, namely  $\varphi(z)= \varphi_bH(z-R_s)$. The convolution Eq.~(\ref{depletion}) is then a measure of the overlap volume --- see Fig.~1. For a separation $l$ between the surface of the large sphere and the wall, the depletion potential reads~\cite{aoJPS58,vrijPAC76}
\begin{equation}
\label{AO}
\beta W^{(0)}(l) = -3\varphi_b \frac{R}{R_s}\left( 1-\frac{l}{2R_s}\right)^2\left(1+\frac{R_s+l}{3R}\right) 
\end{equation}
for $l<2R_s$, and $\beta W^{(0)}(l)=0$ otherwise. The value at contact is $ \beta W^{(0)}(0) = -3\varphi_b \alpha^{-1}  $, where we introduce the aspect ratio $\alpha=R_s/R$. For typical volume fraction $\varphi_b=0.2$ and aspect ratio $\alpha=0.1$, the depth of the potential well is $W^{(0)}(0)=-6k_BT$. Thus, these entropic effects cannot be ignored as they may eventually overcome the direct interactions. Actually, they appear to be essential in describing phase separation and aggregation phenomena in polydisperse mixtures~\cite{russelbook}. 

While much of the theoretical and numerical effort assumes that the small particles are structureless spheres, experimental studies mainly use polymers as depleting agents. Actually, the formal mapping of a polymer solution onto a fluid of ``soft'' particles is a challenging problem because of intricate many-body interactions~\cite{louisJCP01, louisJCP02}. For most of practical purposes however, the exact nature of the small components can be ignored: at low concentrations, polymer coils can be regarded as spherical particles of radius $R_g$, their radius of gyration. Note that this assertion is still valid above the overlap concentration, provided that one substitutes $R_g$ with the correlation length of the solution~\cite{joannyJpolSci79,ohshimaPRL97}.

 In the recent years, a variety of experimental techniques have been developed to probe soft materials at the sub-micrometer scale: optical tweezers, surface force apparatus (SFA), atomic force microscopy (AFM), or total internal reflection microscopy (TIRM).  These progress have enabled direct comparison with theoretical predictions, and numerous experimental works have confirmed the validity of Eq.~(\ref{AO}) for volume fractions $\varphi_b \leq 0.25$~\cite{ohshimaPRL97,kaplanPRL94,rudhartPRL98}. At higher concentrations, a repulsive barrier forms at distances $l\sim 2R_s$, followed by secondary minima reminiscent of the layered structure of the small-component liquid~\cite{richettiPRL92,maoPhysA95, bibenJPCM96, gotzelmannPRE98,crockerPRL99}. Eventually, kinetics effects will prevent the colloidal particle to reach the primary minimum~\cite{andersonNature02}. In the present work, we assume that the packing fraction is low enough to neglect the correlations between small particles. The kinetics is also ignored throughout the treatment~\cite{vliegPeprint02}.

\begin{figure}
\centering
\includegraphics[angle=90]{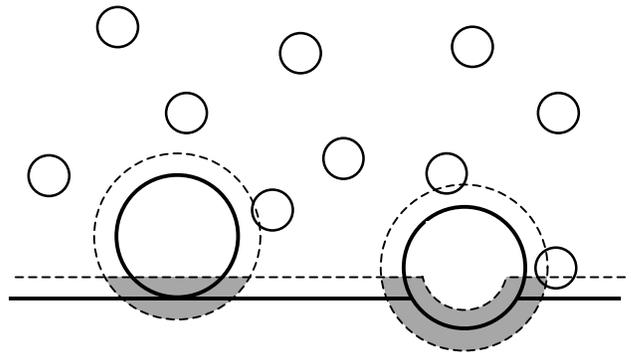} 
\caption{Illustration of the depletion attraction phenomenon. The center-of-mass of the small particles is excluded from a shell of thickness $R_s$ around the large spheres and along the surface (delimited by the doted lines). When these depletion volumes overlap (grey regions), the osmotic pressure of the gas of small particles becomes anisotropic and the large particle is pushed toward the surface. The effective volume available for the small particles is further increased if the surface can deform and envelop the large particle.}
\label{fig1}
\end{figure}

\section{Depletion-induced wrapping \label{sec3}}

The adsorption of colloidal particles on a soft surface is an essential step in many biological processes. The underlying physics of this mechanism has been extensively studied for simple model systems~\cite{dietrichJP297, desernoJPCB02, nogushiBPJ02, boulbitchEPL02}. Here, we investigate the special case of adhesion  driven by depletion forces. We specify the interface profile through the displacement field  $u({\bf r_{\parallel}})$, where ${\bf r_{\parallel}}=(x,y)$ spans the horizontal reference plane at $z=0$, and define the Fourier representation $u_{\bf q}=\int d^2{\bf r_{\parallel}} \exp(-i{\bf qr_{\parallel}})u({\bf r_{\parallel}})$. The statistical behavior of the surface can be studied within a continuum model based on the following Hamiltonian~\cite{helfrichZfP73}
\begin{equation}
\label{hamiltonian}
 \mathcal{H}_\mathcal{I} = \frac{1}{2}\int \frac{d^2{\bf q}}{(2\pi)^2} \vert u_{\bf q} \vert^2 \left(r+\sigma q^2 + \kappa q^4 \right) \ ,
\end{equation}
with $\kappa$ the bending rigidity and $\sigma$ the surface tension. The elastic coefficient $r$ is related to the curvature of an eventual external potential (gravitational field, effective ``mean-field'' intermembrane interactions). This model allows us to describe a wide class of surfaces, ranging from liquid-liquid or liquid-gas interfaces, with high surface tension and vanishing bending modulus, to pure fluid membranes with finite bending modulus and zero surface tension. The intermediate combination where both $\sigma$ and $\kappa$ are finite would for instance correspond to a surfactant monolayer at the oil/water interface~\cite{avibillbook}.

A flexible surface as described by the Hamiltonian $\mathcal{H_I}$ is not a simple planar surface, but i) can be deformed under an external constraint, and ii) presents an intrinsic roughness due to thermal fluctuations. In this section, we derive the phase diagram for the different wrapping states of the colloid. For the sake of simplicity, we restrict ourselves to the bending regime $r=\sigma=0$. The critical behavior of the system results from a balance between the driving depletion force and the membrane bending energy. The former separates into an entropic and an enthalpic contribution, that we study separately.

\subsection{Depletion \emph{vs.} Helfrich repulsion} 

A colloidal particle feels in the vicinity of the membrane a steric repulsion resulting from the thermal undulations~\cite{helfrichZfP78}. An analytical expression for this repulsion is known only for a membrane close to an infinite, flat wall: the free energy at given mean separation $l$ scales as $V_{Helfrich}(l)= A (k_BT)^2/(\kappa l^2)$, with $A=\pi^2/128$~\cite{kleinertPhysLettA99}.  One way to make contact with the curved geometry is to use the Derjaguin approximation~\cite{derjaguinKollZ34}, which relates the force between a sphere and a plane to the potential between two planes. In the limit $l \ll R$, the force exerted by the fluctuating surface on the spherical particle is $F(l)=2\pi R\, V_{Helfrich}(l)$. Here, $l$ is the distance between the surface of the particle and the average position of the membrane. It follows that the effective potential between a sphere and a membrane equals $V_{rep}(l)=\int_l^{\infty}F(z)dz$. For $l<2R_s\ll R$, the total potential  $V(l)  =  V_{rep}(l)+ W^{(0)}(l)$  is
\begin{equation}
\label{potentialtotal}
\beta V(l)  =   \frac{\pi^3}{64 \, \tilde{\kappa}} \left(\frac{R}{l}\right) - 3\varphi_b\alpha^{-1}\left(1-\frac{l}{2R_s}\right)^2   \ ,
\end{equation}
with $\tilde{\kappa}=\kappa/(k_BT)$, whereas only the repulsive part survives at separations $l>2R_s$.
Minimizing $V$ gives the position of the minimum $l_{e} \approx( \pi^3/(192 \, \tilde{\kappa} \, \varphi_b))^{1/2}R_s  $.
 Moreover, the requirement that $l_e$ corresponds to a \emph{global} minimum defines a critical bending modulus 
\begin{equation}
\label{critbend1}
\tilde{\kappa}_{c,1} = \frac{\pi^3}{192 \,  \varphi_b}  \ .
\end{equation}
Below this critical value, there is no bound state and the particle can never reach the surface. In the adhesive regime $\tilde{\kappa}>\tilde{\kappa}_{c,1}$, the particle localizes at the minimum $l_e$ of the potential that equals $l_e \approx 0.5R_s$ for $\tilde{\kappa}=10$ and $\varphi_b=0.2$. 

One could argue that the superposition of the Helfrich repulsion and the direct potential in an attempt to produce an effective potential is in general not correct~\cite{lipowskyPRL86}. Rather, a virial expansion of the direct part is needed to describe the thermodynamics of the system~\cite{milnerJPhysI92}. Here however, both the attractive and repulsive parts of the potential are entropic in nature, thus possibly justifying such a superposition.

\subsection{Depletion \emph{vs.} bending energy}

We now assume that once the repulsive barrier is overcome, the colloidal particle is adsorbed on a purely elastic sheet.  As shown on Fig.~1, the depletion contribution to the free energy is further decreased if the membrane envelops the particle. Note that we assume here that the solvent on both sides of the surface is incompressible: the only volume contribution comes from the variation of the overlap volume. The natural  parameter that describes the wrapping transition is the degree of penetration $\omega$ defined as the ratio between the covered area and the total area of the colloid. The equilibrium value of $\omega$ is set by the balance between osmotic and bending forces. Neglecting the fluctuations, the free energy contributions divides into three parts. The first term is the bending energy of the covered area that simply equals $8\pi\kappa\omega$. In the limit of a small aspect ratio, the overlap volume is $\Delta V_{ov}=8\pi R^3\omega\alpha +\mathcal{O}(\alpha^2)$, so that the depletion-induced adsorption can be described in terms of a constant adhesion energy per unit area $\Gamma = 2k_BT\rho_bR_s$~\cite{desernoJPCB02}. This approximation is certainly not valid in the early stages of the wrapping process where $\omega \sim \alpha$, but remember that we are interested by the limit $\omega \sim 1$. The last contribution arises from the free part of the membrane and is more difficult to evaluate, since it generally requires one to solve the full non-linear shape equations~\cite{seifertPRA90,seifertAdvPhys97,tordeuxPRE02}. Fortunately the problem greatly simplifies for a membrane with no surface tension, providing that there is no pressure difference between both sides of the surface. In this particular case, no forces are applied to the free part of the membrane. The deformation profile becomes a catenoid, a surface with zero curvature and consequently a \emph{vanishing bending energy}. The free energy of the system is then a linear function of $\omega$
\begin{equation}
\label{energybend}
\beta \Delta F(w)=  8\pi \omega  \left(\tilde{\kappa}-\frac{3}{4\pi} \varphi_b \alpha^{-2} \right)   \ .
\end{equation}
Eq.~(\ref{energybend}) then defines a second critical bending rigidity 
\begin{equation}
\label{critbend2}
\tilde{\kappa}_{c,2} =  \frac{3}{4\pi} \varphi_b \alpha^{-2}    \ .
\end{equation}
The colloidal bead is in the unwrapped state above $\tilde{\kappa}_{c,2}$, whereas the membrane completely encapsulates the particle below this critical value~\cite{desernobick02}. It is worthwhile to note that the minimum of the free energy is not an extremum, but corresponds to either boundaries of the domain $\omega =0$ or 1. Therefore, the equilibrium profile does not have to satisfy the usual stationarity requirement $c^{\star}=1/R+\sqrt{2 \Gamma/\kappa}$, with $c^{\star}$ the mean curvature at the detachment point~\cite{seifertPRA90}.

\begin{figure}
\centering
\includegraphics{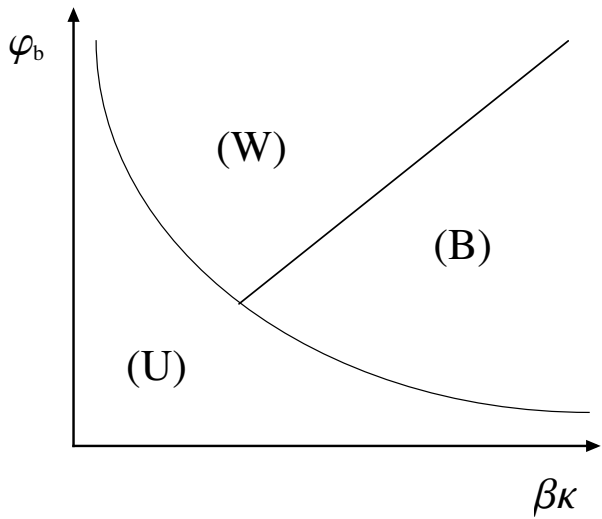} 
\caption{Schematic phase diagram for the wrapping transition. (U): unbound state. (B): bound state. (W): wrapped state.}
\label{fig2}
\end{figure}

The conditions Eq.~(\ref{critbend1}) and (\ref{critbend2}) enable us to draw the phase diagram of the system presented in Fig.~\ref{fig2}. For low bending rigidity or volume fraction, the colloid is always in the unbound state because of the Helfrich repulsion. Also, the membrane cannot wrap the particle if the bending modulus is too high. Conversely, the membrane envelops the colloid when the bending modulus lies within the range $\tilde{\kappa}_{c,1}<\tilde{\kappa}<\tilde{\kappa}_{c,2}$. For typical values $\alpha=0.1$ and $\varphi_b=0.2$ the boundaries correspond to $\tilde{\kappa}_{c,1} \approx 0.77$ and $\tilde{\kappa}_{c,2} \approx4.77$. Depending on the volume fraction and the aspect ratio of the small colloids, this interval can actually vanish: the condition of existence of a wrapped state is given by the inequality 
\begin{equation}
\label{wrapped}
\frac{\alpha \pi^2}{12} <   \varphi_b <1  \ .
\end{equation}

Arguably, the heuristic derivation presented here is not accurate enough to predict the nature of the transitions and the exact values of the numerical constants. The influence of finite-size effects on the unbinding transition  would  in particular deserve a further analysis in the future. However, this phenomenological description provides some important insights for the understanding of the wrapping transition.

\section{Fluctuations of the interface and corrections to the depletion interaction  \label{sec4}}

We derived in the preceding section the phase diagram for the wrapping of a colloidal particle. In particular, we found that encapsulation can indeed be driven by depletion forces. However, we have not yet addressed another relevant question arising from the surface softness, namely the effect of thermal undulations on the depletion potential itself. As explained in Section II, the derivation of this potential only involves the density profile of the small components. Here we adopt a procedure in a sense opposite to the original route pursued by Asakura and Oosawa, and construct the equilibrium concentration profile for a \emph{polymer} solution, instead of spherical beads, near a fluctuating surface. We then apply this result to evaluate the effective depletion potential.

\subsection{Statistical mechanics} 

We consider a generic interface described by the Helfrich Hamiltonian Eq.~(\ref{hamiltonian}). The thermodynamic properties are specified by two lengths $\xi_{\parallel}$ and $\xi_{\perp}$~\cite{seifertAdvPhys97}: $\xi_{\parallel}$ characterizes the exponential decay of the height-height correlation function
\begin{equation}
\label{correl}
G({\bf r_{\parallel}}) = \left\langle   u({\bf r_{\parallel}})u(0)  \right\rangle_\mathcal{I} \sim \, e^{-r_{\parallel}/ \xi_{\parallel}}  \ ,
\end{equation}
whereas $\xi_{\perp}= \left\langle \,   u({\bf r_{\parallel}})^2  \right\rangle^{1/2}_\mathcal{I}$ is a measure of the roughness of the surface.
For instance, a fluid membrane has a vanishing surface tension $\sigma = 0$ and the in and out-of-plane correlations lengths are $\xi_{\parallel}=(4\kappa/r)^{1/4}$ and $\xi_{\perp}=\xi_{\parallel}/(4\sqrt{\tilde{\kappa}})$, respectively.

If a colloidal solution is in contact with this surface, we argue that the rigorous treatment of the  fluctuations should be as follows: assume that the density profile $\rho[u]$ is known for a given conformation $u({\bf r_{\parallel}})$ of the surface. The thermal fluctuations can then be traced out according to 
\begin{equation}
\label{averagerho}
\rho(z)= \left\langle  \rho({\bf r_{\parallel}},z) \right\rangle_\mathcal{I} = \frac{\displaystyle \int \mathcal{D}ue^{-\beta \mathcal{H}_\mathcal{I}}\rho[u]}{\displaystyle \int \mathcal{D}ue^{-\beta \mathcal{H}_\mathcal{I}}}  \ ,
\end{equation}
so that $\rho(z)$ is finally expressed as a functional of the Fourier transform of the surface correlation function
\begin{equation}
\label{fouriercorrel}
G({\bf q}) = \langle \vert u_{\bf q} \vert^2 \rangle_{\mathcal{I} }=\frac{k_BT}{r+\sigma q^2 +\kappa q^4}  \ .
\end{equation}

The remaining problem is to evaluate the functional $\rho[u]$. To this end, we will assume that the depleting particles are linear polymer chains  instead of spherical beads. This idea might at first seem questionable as a macromolecule has many more degrees of freedom than a simple hard sphere. Nevertheless, the partition function $\mathcal{Z}_N({\bf r_{\parallel}},z)$ of an ideal chain of $N$ monomers with one extremity fixed at $({\bf r_{\parallel}},z)$ can be obtained using an analogy to Brownian motion~\cite{doiedwardsbook}. The complexity of surface impenetrability then reduces to a boundary condition problem that  can be solved analytically. The polymer concentration is then expressed in the low-density limit as~\cite{pggbook}
\begin{equation}
\label{rhopolymer}
\rho({\bf r_{\parallel}},z) =  \frac{\rho_b}{N}\int_0^Ndn \mathcal{Z}_n({\bf r_{\parallel}},z)\mathcal{Z}_{N-n}({\bf r_{\parallel}},z) \ .
\end{equation}
Far from the wall, the bulk value equals $\rho_b=\mathcal{N}/V$ with $\mathcal{N}$ the total number of chains in solution. The average value Eq.~(\ref{averagerho}) of the density profile  follows immediately. 

The evaluation of $\mathcal{Z}_N$ for a any surface profile $u({\bf r_{\parallel}})$ is in general a fairly complex task~\cite{honeMacro87,hankePRE99}. Here, we follow a perturbative approach based on the work of polymers grafted to a fluid membrane published elsewhere~\cite{bickelPRE00,bickelEPJE01}. The key ingredient of the derivation is to write the partition function as a series $\mathcal{Z}_N=\sum_i \mathcal{Z}_N^{(i)}$, where $\mathcal{Z}_N^{(i)}$ is of order $u^i$. The expansion of the partition function of a Gaussian chain in the vicinity of a non-planar surface is detailed in the Appendix.  All the results displayed in the following are exact up to linear order in the natural expansion parameter $\tilde{\kappa}^{-1}$. 

At this point, it should be mentioned that our perturbative scheme is only valid in the limit of ``almost flat surface'', i.e.,  for $\xi_{\perp}$ much smaller than the radius of gyration $R_g$ of the coils. This condition can be written in terms of the in-plane correlation length as
\begin{equation}
\label{condition}
\left(\frac{\xi_{\parallel}}{R_g}\right)^2 \ll 16 \tilde{\kappa}  \ .
\end{equation}
It implies that our calculations are valid even for $R_g<\xi_{\parallel}$, provided that the bending rigidity is large enough. For instance, for a bending modulus $\tilde{\kappa}=10$, the consistency of the approach is still fulfilled for $\xi_{\parallel}$ as high as $\xi_{\parallel}=10R_g$.

\subsection{Gibbs adsorption}

We begin the discussion with the first moment of the distribution, which is less computationally demanding. In the vicinity of the interface, the proportion of polymers that belongs to the adsorption or depletion layer may be quantified by the adsorbance $\Gamma$ defined as~\cite{widombook}
\begin{equation}
\label{adsorbance}
\Gamma=\int_0^{\infty} dz \left(\rho(z)-\rho_b \right) \ .
\end{equation}
The adsorbance usually scales as $\Gamma \sim \xi \rho_b$, with $\xi$ the typical length-scale over which the concentration relaxes towards the bulk value. Note that the length $\xi$ is an algebraic quantity: a positive value  reflects particle accumulation, whereas a negative value expresses a deficit of particles in a layer of thickness $\xi$. For instance, the concentration profile $\rho^{(0)}$ of ideal chains depleted from a flat, infinitely repulsive surface  is~\cite{marquesMacromol90,eisenbook} 
\begin{equation}
\label{densityzero}
\begin{split}
\rho^{(0)}(\zeta)=  &  \rho_b
 \Big\{   \zeta^2\Big[1+ \erf(\zeta/2)-2\,\erf(\zeta)\Big] + 2\,\erf(\zeta/2) \\
&    
   -\,\erf(\zeta) -\frac{2}{\sqrt{\pi}}\left[e^{-\zeta^2}-e^{-\zeta^2/4}\right] \Big\}  \ ,
\end{split}
\end{equation}
with $\zeta=z/R_g$ the (dimensionless) distance from the wall, and $R_g=\sqrt{Na^2/6}$ the radius of gyration of the polymer coil. The corresponding adsorbance is~\cite{tuinierJCP00} 
\begin{equation}
\label{adsorbancezero}
\Gamma^{(0)}=-\frac{2}{\sqrt{\pi}} \rho_b R_g  \ .
\end{equation}

In the rigid wall limit ($\tilde{\kappa}\rightarrow \infty$), the particles are excluded from the vicinity of the surface over a distance $d_{\infty}  \sim R_g$. If the prescription $R_s=2R_g/\sqrt{\pi}$ is used, the excess surface density for the hard sphere and the polymer solutions are the identical, which illustrates the connection with the original model of Asakura and Oosawa.

For a finite rigidity ($\tilde{\kappa} < \infty$), one would at first expect the thickness of the depletion layer to increase according to $d_{\tilde{\kappa}} \sim  R_g +\xi_{\perp} $. This can be checked by  performing the trace over the surface conformations Eq.~(\ref{averagerho}).  We find the correction $\Delta \Gamma = \Gamma -\Gamma^{(0)}$ to the adsorbance 
\begin{equation}
\label{deltadsorbance}
\begin{split}
 \Delta \Gamma  =\displaystyle{ -\frac{1}{\sqrt{\pi}} \frac{\rho_b}{ R_g } \int  \frac{d^2{\bf q}}{(2\pi)^2}}  
&
 G({\bf q})   \Big(e^{-q^2R_g^2}  
 \\ & 
+\sqrt{\pi}qR_g\erf(qR_g)-1 \Big) \ ,
\end{split}
\end{equation}
where the correlation function $G({\bf q})$ is defined by Eq.~(\ref{fouriercorrel}). One should note that $\Delta \Gamma$ is always negative, reflecting an \emph{expansion} of the depletion layer. Expression~(\ref{deltadsorbance}) can be evaluated for a membrane without surface tension: we get for $\sigma=0$
\begin{equation}
\label{asympadsorbance}
\Delta \Gamma \sim
\begin{cases}
\displaystyle -\tilde{\kappa}^{-1}  \rho_b \xi_{\parallel}  & \text{for }   \xi_{\parallel} \ll R_g,  \\
\displaystyle -\tilde{\kappa}^{-1} \rho_b R_g   & \text{for }   \xi_{\parallel} \gg R_g.
\end{cases}
\end{equation}
We see that in both limits, the depletion layer is increased by $\tilde{\kappa}^{-1}\min(\xi_{\parallel},R_g)$. This seems to confirm the ``naive'' expectation that arises from the scaling argument. Nevertheless, this picture happens not to be completely accurate, since the adsorbance is only the first moment of the distribution and therefore does not contains all the information about the density profile.

\subsection{Correction to the density profile}

More difficult to evaluate analytically is the correction $\Delta \rho (z)= \rho (z)-\rho^{(0)}(z)$ to the concentration profile. We first focus on the proximal region $z \ll R_g$~\cite{remark3}. The value at the origin is easily obtained and equals
\begin{equation}
\label{profileorigin}
\Delta \rho (0) = \frac{ \rho_b \, }{16\tilde{\kappa}} \left(\frac{\xi_{\parallel}}{R_g}\right)^2  \ .
\end{equation}
Quite unexpectedly, we find that the polymers \emph{accumulate} near the surface: $\Delta \rho (0)>0$. Moreover, if we extend the calculations up to linear order in $z$ we obtain for a generic surface
\begin{equation}
\label{profilelinear}
\begin{split}
\Delta \rho (z) = \Delta \rho (0)- & \frac{\rho_b}{\sqrt{\pi} }\frac{z}{R_g^3 }
  \int \frac{d^2{\bf q}}{(2\pi)^2}  G({\bf q})\Big(4e^{-q^2R_g^2} \\
  &
 +4\sqrt{\pi}qR_g\erf(qR_g)-1 \Big)   \ .
\end{split}
\end{equation}
Since the integrand is positive for any value of the parameters, the correction to the depletion profile starts with a \emph{negative} slope. For a membrane without surface tension, we get the following limiting cases
\begin{equation}
\label{asympslope}
\Delta \rho(z) \simeq
\begin{cases}
\displaystyle \Delta \rho (0)\Big( 1-\frac{2\sqrt{2}z}{\xi_{\parallel}}\Big)  & \text{for }  R_g \gg \xi_{\parallel} ,  \\
\displaystyle \Delta \rho (0)\Big( 1-\frac{3z}{\sqrt{\pi}R_g}\Big)  & \text{for }  R_g \ll \xi_{\parallel} .
\end{cases}
\end{equation}

\begin{figure}
\centering
\includegraphics{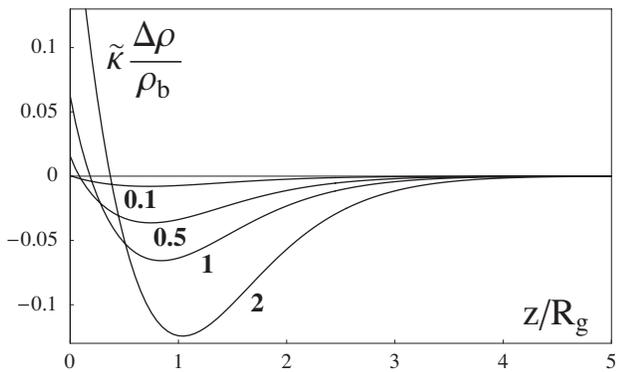}
\caption{Correction to the density profile for an interface with vanishing surface tension ($\sigma=0$). The different curves correspond to $\xi_{\parallel}/R_g=0.1,0.5,1 \ \text{and }2$, respectively.  In the proximal region, membrane fluctuations favor polymer adsorption on the surface; at larger distances, the chains are pushed back into the bulk by the Helfrich repulsion.}
\label{fig3}
\end{figure}

This shows that the simple picture emerging from the scaling argument is in fact erroneous. Formally, the behavior described by Eqs.~(\ref{profileorigin})--(\ref{asympslope}) is reminiscent to the adsorption of ideal polymer chains on a flat wall~\cite{pggbook}. It should be emphasized that the effects discussed here are more than a simple correction to $\rho^{(0)}$. Indeed, expanding Eq.~(\ref{densityzero}) for $z \ll R_g$ gives $\rho^{(0)}(z)\propto z^2$, which vanishes at the surface. Therefore, Eq.~(\ref{profilelinear}) specifies not only the short-distance behavior of the correction $\Delta \rho$ but actually describes the \emph{total} concentration profile $\rho= \rho^{(0)}+\Delta \rho$. It is even possible to achieve a density at $z=0$ comparable to the bulk density, but for values of the parameters that reach the limit of validity of our approach --- see Eq.~(\ref{condition}).

\begin{figure}
\centering
\includegraphics{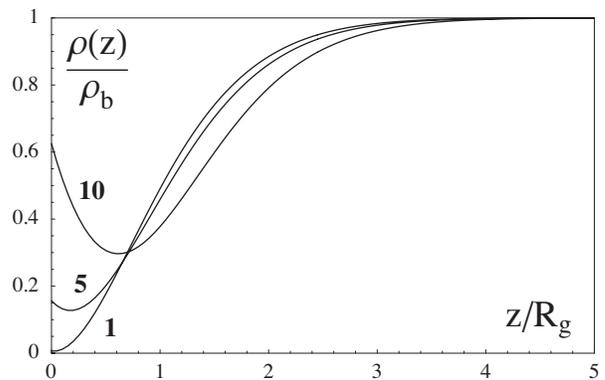}
\caption{Complete density profile $\rho=\rho^{(0)} + \Delta \rho$ for an interface with $\sigma=0$ and $\tilde{\kappa}=10$. The different curves correspond to $\xi_{\parallel}/R_g=1,5 \ \text{and }10$, respectively.}
\label{fig4}
\end{figure}

The analytical expression for $\Delta \rho$ is uninstructive and will not be given here. Fig.~\ref{fig3} presents the correction to the density profile for different values of the ratio $\xi_{\parallel}/R_g$~\cite{remark}. As previously noted, the coils are first adsorbed in the proximal region, i.e., the chains have the tendency to ``fill'' the thermal undulations. Interestingly, the chains are \emph{repelled} more from the fluctuating surface at intermediate distances. This effect ultimately dominates the overall profile since the adsorbance Eq.~(\ref{deltadsorbance}) is negative. Finally, the interface affects the distribution over a finite extension $\sim 4R_g$.
As illustrated in Fig.~\ref{fig3}, the effect is rather minor for small values of the ratio $\xi_{\parallel}/R_g$. On the other hand, the correction cannot be ignored for $\xi_{\parallel}>R_g$, and we plot in Fig.~\ref{fig4} the full depletion profile $\rho(z)=\rho^{(0)}(z)+\Delta \rho(z)$ for a bending rigidity $\tilde{\kappa}=10$.

\subsection{Effective depletion potential}

\begin{figure}
\centering
\includegraphics{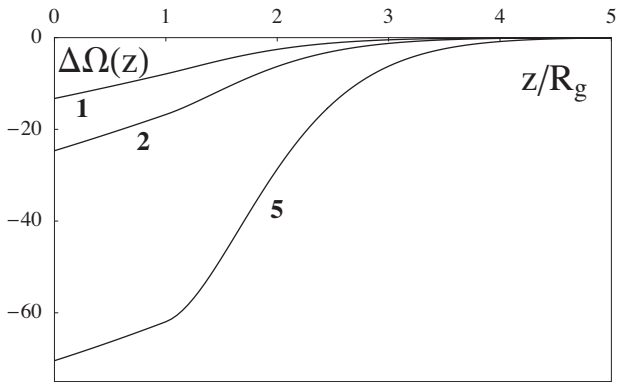}
\caption{Correction to the depletion potential in dimensionless units: $\Delta \Omega = (\tilde{\kappa}/\rho_bR_g^3)\Delta W$, for $\xi_{\parallel}/R_g=1,2  \ \text{and }5$ respectively. The inflection point is a manifestation of the non-monotonic behavior of $\Delta \rho(z)$.}
\label{fig5}
\end{figure}

The non-monotonic behavior revealed in the preceding section strongly affects the depletion potential between the colloidal particle and the surface, mediated by the small polymer coils. At this point of the derivation, we have to evaluate  the convolution product Eq.~(\ref{depletion}) numerically since it involves the complete density profile. For the sake of simplicity, we assume that the Mayer function characterizing the polymer/colloid interactions is still a step function. The correction $\Delta W=W-W^{(0)}$ is depicted in Fig.~\ref{fig5} in reduced units, where the combination $\rho_b R_g^3$ is analogous to the volume fraction $\varphi_b$ of spherical particles. For reasonable values $R=10R_g$, $\rho_bR_g^3=0.2$, $\tilde{\kappa}=10$ and $\xi_{\parallel}=5R_g$, the depth of the well is increased by a substantial amount $\Delta W(0)\approx -1.4k_BT$. In terms of the Asakura-Oosawa potential Eq.~(\ref{AO}), this represents 25\% of the original value at the wall. Of course, one has certainly to take into account the \emph{direct} interaction between the colloidal particle and the surface at distances smaller than $\xi_{\perp}$, but this question is beyond the scope of this report. For $\xi_{\parallel}=5R_g$ and $\tilde{\kappa}=10$, this distance is approximatively $\xi_{\perp}\approx 0.4R_g$. Note also that the curves present an inflection point at a separation matching the minimum of $\Delta \rho (z)$. As $\xi_{\parallel}$ increases, this point eventually becomes an minimum of $\Delta W(z)$. 

Actually, the amplitude of the effective potential can be inferred from very general arguments. Let us start with a solution of particles confined between two planes. It is easy to show~\cite{remark2} that the value $w(0)$  (per unit area) of the depletion potential at contact  scales as $\beta w(0)\sim \Gamma^{(0)}$, with the adsorbance given by Eq.~(\ref{adsorbancezero}). In the Derjaguin approximation, the corresponding force between a spherical particle of radius $R$ and a plane is $F\sim  w(0)R$. Since the range of the depletion interaction is fixed by the size $R_s$ of the depleting particles, the value at contact of the depletion potential between a colloidal particle and a plane is $\beta W^{(0)}(0) \sim  R R_s \Gamma^{(0)}$, in perfect agreement with Eq.~(\ref{AO}). Assuming the same scaling for the correction to the potential, we expect $\Delta W$ to behave like $\beta \Delta W(0) \sim  R R_s \Delta \Gamma$. It is more instructive to consider the ratio $\Delta W/W^{(0)}$
\begin{equation}
\label{asympratio}
\frac{\Delta W(0)}{W^{(0)}(0)} \sim
\begin{cases}
\displaystyle \tilde{\kappa}^{-1} \left(\frac{\xi_{\parallel}}{R_g}\right) & \text{for }   \xi_{\parallel} \ll R_g ,  \\
\displaystyle \tilde{\kappa}^{-1} & \text{for } \xi_{\parallel}  \sim R_g ,
\end{cases}
\end{equation}
showing that the effects discussed here can be more than a minor correction.

\section{Discussion  \label{sec5}}

The simple analysis described in this report has provided some very interesting and somewhat unexpected results for depletion effects at fluctuating interfaces. The phenomenological approach initiated in the first part of this study  reveals that entropic interactions can indeed lead to full particle coverage. This route to particle encapsulation is particularly attractive since it benefits from the wide versatility offered by depletion forces.  

The main part of this work is dedicated to the concentration profile of small particles near a fluctuating surface. For the last decade, intensive studies have focused on the way a colloidal suspension affects the elastic properties of the surface. Particularly well documented is the adsorption of polymers from dilute or semi-dilute solutions~\cite{brooksJPhysII91, clementJPhysII97, skauEPJE02}, as well as the depletion of diverse objects like rigid rods~\cite{yamanPRL97}, ideal or self-avoiding polymers~\cite{hankePRE99,louisbJCP02,bickelEPJE02}. To the best of our knowledge,  it is the first time that the opposite question, namely, the effect of the surface on the bulk properties, is theoretically addressed. For this derivation we make three major hypothesis: i) the particles are ideal polymer chains, ii) we treat only the very dilute case where the polymers act independently, and iii) we assume that surface undulations are only a perturbation to the planar configuration. Even within this restrictive set of assumptions, we find the following remarkable results. In the first regime $z<\xi_{\perp}$, the polymers are closely bound to the surface by an effective, attractive potential. This primarily results from the tendency of the chains to fill the holes or valleys on a rough surface, and can be understood as follows: close to a wall, the colloidal solution is affected over a characteristic length fixed by the size $R_s$ of the particles. This naturally gives rise to a surface tension, defined as the free energy penalty per unit area to create the depletion layer. If the wall is slightly curved with radius $R$, the excess surface energy $\gamma_s$ can be expanded in powers of the ratio $q=R_s/R$ 
\begin{equation}
\label{tension}
\gamma_s(q)=\gamma^{(0)}\left( 1+\alpha q +\beta q^2 \right) + \mathcal{O}(q^3)  \ ,
\end{equation}
with $\gamma^{(0)}$ the planar wall contribution that scales as $\gamma^{(0)} \sim k_BT\rho_bR_s$. Note that with this convention, a sphere has a positive curvature. The first correction $\alpha$ controls the affinity of the solution for the curved surface. It can be shown for ideal polymers that $\alpha>0$, meaning that the surface energy is lowered when the chains are in the vicinity of a concave region ($R<0$). In the language of curvature energy, one would infer that the surface bends more favorably \emph{toward} the polymers with a spontaneous radius of curvature $R_0\approx-R_g$. In the reverse problem considered here we can conclude that the chains have a higher affinity for the concave regions, resulting in a \emph{finite} concentration at $z=0$ after averaging over the surface degrees of freedom. This effect becomes more pronounced as one increases the in-plane correlation lengths. Indeed, the polymer coils can never fit into the fluctuations when $\xi_{\parallel} \ll R_g$ and the correction is negligible. This is not the case in the regime where $\xi_{\parallel} \gtrsim R_g$. Here, the statistical weight of the undulations with curvature at their minima smaller than $1/R_g$ grows rapidly, followed by an increase of $\Delta \rho (0)$ as shown in Fig.~\ref{fig3}.

In the range $\xi_{\perp}<z<4R_g$, the chains are on the contrary repelled more from the fluctuating surface than from a flat wall. This unusual behavior reflects a broadening of the interface: as the polymer coils move away from the average elevation $z=0$, they ``feel'' an effective layer of thickness $\xi_{\perp}$ from which they are excluded. This repulsion actually corresponds to the force exerted by the fluctuating surface on a particle, and suggests that finite-size effects may lead to drastic changes to the Helfrich repulsion. This question is part of ongoing investigations and will be the subject of a future publication. 

As a final remark, we point out that our formulation describes the particles as ideal chains. It is therefore natural to wonder how the results may be generalized to other species. A partial answer can be built following the ``surface tension'' argument presented below. It is known that the first order correction in Eq.~(\ref{tension}) for self-avoiding walks and spherical beads is comparable to that found for ideal chains~\cite{louisbJCP02}, so that it is tempting to extend our conclusions concerning particle accumulation at the wall. This is not completely obvious however as the second-order coefficient $\beta$ in the development Eq.~(\ref{tension}) does not follow this trend: $\beta$ is \emph{positive} for hard spheres, \emph{negative} for self-avoiding walks, and \emph{zero} for ideal polymers. Nevertheless, the physical picture presented in this report is expected, at least qualitatively, to remain valid, and will await for further numerical and experimental verification.  

\acknowledgments

M. Deserno, W. Gelbart, and D. Roux are deeply acknowledged for enjoyable and valuable discussions. This work has also beneficiated from the comments and encouragments of R. Bruinsma and C. Marques. Lastly, the author is particularly indebted to E. Torres for a critical reading of the manuscript.

\appendix*

\section{Perturbative calculation of $\mathcal{Z}_N$}

The fundamental object in polymer statistic is the Green function or propagator $G_N(\{{\bf r_{\parallel}},z\},\{{\bf r_{\parallel}'},z'\}) $ of the chain. It represents the statistical weight of the random walk that starts at $\{{\bf r_{\parallel}},z\}$ and ends at $\{{\bf r_{\parallel}'},z'\}$ in $N$ steps. The Green function solves the Edwards equation~\cite{doiedwardsbook}
\begin{equation}
\label{edwardsgreen}
\left( \frac{\partial}{\partial  n}-b^2\nabla^2\right)G_{n}=\delta(n)\delta({\bf r_{\parallel}}-{\bf r_{\parallel}'})\delta(z-z')  \ ,
\end{equation}
with $b=a/\sqrt{6}$, $a$ being the monomer size. The surface impenetrability is expressed as a boundary condition $G_N(\{{\bf r_{\parallel}},u({\bf r_{\parallel}})\},\{{\bf r_{\parallel}'},z'\}) =0$.  
For the flat, reference case, the Green function can be factorized
\begin{equation}
\label{edwardsgreen1}
\begin{split}
G_N^{(0)}&(\{{\bf r_{\parallel}},z\},
 \{{\bf r_{\parallel}'},z'\})= \left(4\pi R_g^2\right)^{-3/2} \exp\left[-\frac{({\bf r_{\parallel}}-{\bf r_{\parallel}'})^2}{4R_g^2}\right] \\
& \times \left\{ \exp\left[-\frac{(z-z')^2}{4R_g^2}\right]-\exp\left[-\frac{(z+z')^2}{4R_g^2}\right]\right\} 
\end{split}
\end{equation}
with $R_g=\sqrt{Nb^2}$ the radius of gyration.

The partition function of a chain with one extremity fixed at $\{{\bf r_{\parallel}},z\}$    is defined as $\mathcal{Z}_N({\bf r_{\parallel}},z)=\int d{\bf r_{\parallel}'}dz'G_N(\{{\bf r_{\parallel}},z\},\{{\bf r_{\parallel}'},z'\})$, the integral running over all the space available for the free extremity. For instance, the partition function of a chain with one extremity fixed at distance $z$ from a flat wall is $\mathcal{Z}_N^{(0)}=\erf(z/2R_g)$, where $\erf(x)=\int_0^x du\exp(-u^2)$ is the error function.

The starting point of the derivation is to write the partition function as a series $\mathcal{Z}_N=\sum_i \mathcal{Z}_N^{(i)}$, where $\mathcal{Z}_N^{(i)}$ is of order $u^i$~\cite{bickelPRE00,bickelEPJE01,goldsteinPRA90}. The linearity of Eq.~(\ref{edwardsgreen}) then implies that each term satisfies an Edwards equation individually 
\begin{equation}
\label{edwards}
\left( \frac{\partial}{\partial n}-b^2\nabla^2\right)\mathcal{Z}^{(i)}_n({\bf r_{\parallel}},z)=0  \ .
\end{equation}
In addition, $\mathcal{Z}_N$ verifies the boundary condition $\mathcal{Z}_N({\bf r_{\parallel}},u({\bf r_{\parallel}}))=0$ that we also expand as a series of $u^i$
\begin{equation}
\label{seriebc}
\begin{split}
0 =& \mathcal{Z}_N({\bf r_{\parallel}},u({\bf r_{\parallel}})) \\
=& \mathcal{Z}_N({\bf r_{\parallel}},0)   +   u({\bf r_{\parallel}})  \left. \frac{\partial \mathcal{Z}_N}{\partial z}\right\vert_0   + \frac{1}{2}u({\bf r_{\parallel}})^2  \left. \frac{\partial^2 \mathcal{Z}_N}{\partial z^2}\right\vert_0  +  \ldots 
\end{split}
\end{equation}
The problem is then solved recursively. The idea underlying this derivation is to recast the problem of modulated boundary conditions into a series of simpler problems in planar geometry. The calculations are performed up to second order, and we identify the corresponding conditions
\begin{align}
\label{orderbc}
\mathcal{Z}_N^{(1)}({\bf r_{\parallel}},0) & =   -u({\bf r_{\parallel}}) \left. \frac{\partial \mathcal{Z}_N^{(0)}}{\partial z}\right\vert_0  \\
\mathcal{Z}_N^{(2)}({\bf r_{\parallel}},0) & =   -u({\bf r_{\parallel}})  \left.\frac{\partial \mathcal{Z}_N^{(1)}}{\partial z}\right\vert_0 - \frac{u({\bf r_{\parallel}})^2}{2}  \left. \frac{\partial^2 \mathcal{Z}_N^{(0)}}{\partial z^2}\right\vert_0 
\end{align}

The solutions of Eq.~(\ref{edwards}) can then be formally expressed in terms of the Green function $G^{(0)}$ of the problem~\cite{carslawbook}
\begin{equation}
\label{magicfl}
\mathcal{Z}^{(i)}({\bf r_{\parallel}},z) = b^2 \int_0^N dn \int d^2{\bf r_{\parallel}'} \,  \frac{\partial G_n^{(0)}}{\partial z'} \bigg\vert_0 \mathcal{Z}_{N-n}^{(i)}({\bf r_{\parallel}'},0)  \ .
\end{equation}
This convolution product is conveniently written in Fourier-Laplace representation as 
\begin{equation}
\label{magicfl}
\tilde{\mathcal{Z}}^{(i)}({\bf q},z) = b^2 \frac{\partial \tilde{G}^{(0)}}{\partial z'} ({\bf q},z,0)\tilde{\mathcal{Z}}^{(i)}({\bf q},0)  \ ,
\end{equation}
where the Laplace transform $\tilde{F}$ of a function $F_N$ is defined as $\tilde{F}=\int_0^{\infty} dn e^{-sn}F_n$.
It is then fairly simple to obtain the partition function up to second order
\begin{align}
\tilde{\mathcal{Z}}^{(1)}({\bf q},z)  & =    -u_{\bf q}\frac{1}{b\sqrt{s}}\exp\left[-\frac{z}{b}\sqrt{s+(qb)^2}\right] \label{orderfl2} \\
\tilde{\mathcal{Z}}^{(2)}({\bf q},z)  & =   -\frac{1}{2b^2}\exp\left[-\frac{z}{b}\sqrt{s+(qb)^2}\right] \nonumber \\
& \times
  \int \frac{d^2{\bf q'}}{(2\pi)^2}u_{\bf q'}u_{\bf q-q'}  \left\{ 2\frac{\sqrt{s+(q'b)^2}}{\sqrt{s}}-1\right\} \label{orderfl3}
\end{align}
The set of Equations~(\ref{orderfl2}), (\ref{orderfl3}) provides the exact expansion of the partition function of a Gaussian chain in the vicinity of a non-planar surface. The average value Eq.~(\ref{averagerho}) of the density profile defined by Eq.~(\ref{rhopolymer}) follows immediately.


\begin{thebibliography}{99}
\bibitem{remark3}
The term ``proximal'' has been introduced  by de Gennes in the context of polymer adsorption, in order to describe the particular region where the microscopic details of the surface are relevant~\cite{pggbook}. Here however, we use this term with the general meaning of proximity.
\bibitem{remark}
Athough the density is certainly non-zero for slightly negative values of $z$, the profiles are known only for $z \geq 0$. This restriction is actually inherent to the perturbative scheme followed in this work. Note however that the correction to the Gibbs adsorbance takes into account the amount of chains located at $z <0$.
\end{thebibliography}
\end{document}